\documentclass[runningheads]{llncs}
\usepackage{calgo}
\usepackage{version} 
\usepackage[hidelinks]{hyperref}
\hypersetup{colorlinks=true,linkcolor=blue,citecolor=blue,urlcolor=blue}
\usepackage{orcidlink}
\title{On the Formalization of \\ the Notion of a Concurrent Algorithm}
\author{C.A. Middelburg\,\orcidlink{0000-0002-8725-0197}}
\institute{Informatics Institute, Faculty of Science, University of
           Amsterdam \\
           Science Park~900, 1098~XH Amsterdam, the Netherlands \\
           \email{C.A.Middelburg@uva.nl}}
  
\titlerunning
 {On the Formalization of the Notion of a Concurrent Algorithm}
\authorrunning
 {C.A. Middelburg}

\begin{document}
\maketitle

\begin{abstract}
Previous papers give accounts of quests for satisfactory formalizations 
of the classical informal notion of an algorithm and the contem\-porary 
informal notion of an interactive algoritm.
In this paper, an attempt is made to generalize the results of the 
former quest to the contemporary informal notion of a concurrent 
algorithm.
The notion of a concurrent proto-algorithm is introduced. 
The thought is that concurrent algorithms are equivalence classes
of concurrent proto-algorithms under an appropriate equivalence 
relation.
Three equivalence relations are defined.
Two of them are deemed to be bounds for an appropriate equivalence 
relation and the third is likely an appropriate one.
The connection between concurrency and non-deter\-minism in the 
presented setting is also addressed. 

\begin{classcode}
F.1.1, F.1.2, F.2.0 
\end{classcode}
\end{abstract}

\section{Introduction}
\label{sect-intro}

In~\cite{Mid24a} is reported on a quest for a satisfactory formalization 
of the classical informal notion of an algorithm.
By this is meant the notion of an algorithm that is informally 
characterized in many works from the mathematical and computer science 
literature, including standard works such 
as~\cite{Kle67a,Knu97a,Mal70a,Rog67a}.
In the works concerned, an algorithm is informally characterized by 
properties that are considered the most important ones of an algorithm.
The notion of a proto-algorithm defined in~\cite{Mid24a} captures 
virtually all properties that are mentioned in those works.
Moreover, a proto-algorithm expresses a pattern of behaviour to solve 
all instances of a computational problem without using a machine model 
or an algorithmic language.
This makes it neither too abstract nor too concrete to be a suitable 
basis for investigating what exactly an algorithm is in the setting of 
the kinds of computation that are based on a model of parallel 
computation.

In this paper, the algorithms of the kind that emerged due to the advent 
of parallel computation are called concurrent algorithms.
They can be briefly described as follows: ``The pattern of behaviour 
expressed by a concurrent algorithm consists of multiple parts that can 
take place concurrently''.
Algorithms that are called parallel algorithms in the computer science 
literature are usually concurrent algorithms satisfying restrictions 
imposed by some model of parallel computation (cf.~\cite{KRS90a}).
The notion of a concurrent algorithm is intended to encompass all types 
of parallel algorithms that can be found in the computer science 
literature.
Using the term concurrent algorithm avoids confusion with the existing 
uses of the term parallel algorithm.

Despite the fact that parallel algorithms of various types are 
widespread today, no work can be found in the computer science 
literature that aims at a satisfactory formalization of one of the 
informal notions of a parallel algorithm that exist today.
This motivated me to start a quest for a satisfactory formalization of 
the notion of a concurrent algorithm.
An important goal of this quest is gaining more insight into what the 
different types of parallel algorithms have in common and how they all 
differ from classical algorithms.

Due to the advent of parallel computation, several related models of 
parallel computation have been proposed.
Some of them are based on a particular variant of Turing machines, such 
as the parallel Turing machine model of computation proposed 
in~\cite{Wie84a}.
However, most are based on a variant of a RAM (Random Access Machine). 
They include synchronous parallel RAM models of computation, such as the 
models proposed in~\cite{FW78a,Gol82a,SV84a}, and asynchronous parallel 
RAM models of computation, such as the models 
proposed in~\cite{CZ89a,KRS90a,Nis94a}.
In this paper, the starting point for the formalization of the notion of 
a concurrent algorithm is a characterization of the notion by properties 
suggested by those models of parallel computation.

Based on that characterization, the notion of a concurrent 
proto-algorithm is introduced.
Like the notion of a proto-algorithm, the notion of a concurrent 
proto-algorithm is introduced with the thought that concurrent 
algorithms are equivalence classes of concurrent proto-algorithms under 
an appropriate equivalence relation.
Two equivalence relations are defined that give bounds for an 
appropriate equivalence relation and one equivalence relation is defined 
that is likely an appropriate one.
The connection between concurrency and non-determinism in the setting 
of concurrent proto-algorithms and the last equivalence relation is also 
addressed. 

This paper is organized as follows.
First, the basic notions and notations used in this paper are 
introduced (Section~\ref{sect-preliminaries}).
Next, an intuitive characterization of the notion of a concurrent 
algorithm is given by properties that are considered to belong to the 
most important ones of a concurrent algorithm 
(Section~\ref{sect-algo-informal}).
After that, the formal notion of a concurrent proto-algorithm is 
introduced and the isomorphism relation on concurrent proto-algorithms 
is defined (Section~\ref{sect-proto-algo}).
Then, it is defined what a run of a concurrent proto-algorithm 
is and what is computed by a concurrent proto-algorithm
(Section~\ref{sect-step-run-comp}).
Thereafter, two equivalence relations on concurrent proto-algorithms are 
defined (Section~\ref{sect-equivalence}).
Following that, the connection between non-determinism and concurrency 
in the presented setting is addressed 
(Section~\ref{sect-non-det-concur}).
Finally, some concluding remarks are made 
(Section~\ref{sect-conclusions}).

In this paper, an attempt is made to generalize the work on the 
classical notion of an algorithm presented in~\cite{Mid24a} to the 
notion of a concurrent algorithm.
In Sections~\ref{sect-preliminaries}--\ref{sect-proto-algo}, this has 
led to some text overlap with~\cite{Mid24a}.

\section{Preliminaries}
\label{sect-preliminaries}

In this section, the basic notions and notations used in this paper are 
introduced.

The notion of a concurrent proto-algorithm will be formally defined 
in Section~\ref{sect-proto-algo} in terms of three auxiliary notions. 
The definition of one of these auxiliary notions is based on the 
well-known notion of a rooted labeled directed graph.
However, the definitions of this notion given in the mathematical and 
computer science literature vary. 
Therefore, the definition that is used in this paper is given.

\begin{udef}
A \emph{rooted labeled directed graph} $G$ is a sextuple 
$(V,E,\LBLv,\LBLe,l,r)$, where:
\begin{itemize}
\item
$V$ is a non-empty finite set, whose members are called the 
\emph{vertices} of $G$;
\item
$E$ is a subset of $V \x V$, whose members are called the 
\emph{edges} of $G$;
\item
$\LBLv$ is a countable set, whose members are called the 
\emph{vertex labels} of $G$; 
\item
$\LBLe$ is a countable set, whose members are called the 
\emph{edge labels} of $G$; 
\item
$l$ is a partial function from $V \union E$ to $\LBLv \union \LBLe$ 
such that
\begin{itemize}
\item[]
for all $v \in V$ for which $l(v)$ is defined, $l(v) \in \LBLv$ and
\item[]
for all $e \in E$ for which $l(e)$\, is defined, $l(e) \in \LBLe$,
\end{itemize}
called the \emph{labeling function} of $G$;
\item 
$r \in V$, called the \emph{root} of $G$.
\end{itemize}
\end{udef}
The additional graph theoretical notions defined below are also used.
\begin{udef}
Let $G = (V,E,\LBLv,\LBLe,l,r)$ be a rooted labeled directed graph.
Then a \emph{cycle} in $G$ is a sequence 
$v_1\, \ldots\, v_{n+1} \in V^*$
such that, for all $i \in \set{1,\ldots,n}$, $(v_i,v_{i+1}) \in E$,
$\mathrm{card}(\set{v_1,\ldots,v_n}) = n$, and $v_1 = v_{n+1}$.
Let, moreover, $v \in V$.
Then 
the \emph{indegree} of $v$, written $\indeg(v)$, is  
$\mathrm{card}(\set{v' \where (v',v) \in E})$ 
and
the \emph{outdegree} of $v$, written $\outdeg(v)$, is  
$\mathrm{card}(\set{v' \where (v,v') \in E})$. 
\end{udef}

For convenience, the following notions concerning tuples are used:
\begin{udef}
\sloppy
Let $\cA_1,\ldots,\cA_n$ be sets,
let $a_1 \in \cA_1$, \ldots, $a_n \in \cA_n$,
let 
$\boldsymbol{a} = (a_1,\ldots,a_n)$, and
let $i \in \set{1,\ldots,n}$.
Then the \emph{$i$th element of} $\boldsymbol{a}$, written 
$\boldsymbol{a}(i)$, is $a_i$.
Let, moreover $a \in \cA_i$.
Then $\boldsymbol{a}$ \emph{with the $i$th element replaced by} $a$,
written $\updtup{\boldsymbol{a}}{i}{a}$, is
$(a_1,\ldots,a_{i-1},a,a_{i+1},\ldots,a_n)$.
\end{udef}

The following notations may not be entirely standard:
\begin{itemize}
\item
we write $\Natpos$ for the set $\set{n \in \Nat \where n > 0}$ of 
positive natural numbers;
\item
we write $\cA^n$, where $\cA$ is a set and $n \in \Natpos$, for the 
$n$-fold Cartesian product of $\cA$ with itself;
\item
we write $\cP(\cA)$, where $\cA$ is a set, for the set of all subsets of 
$\cA$.
\end{itemize}

In Section~\ref{sect-proto-algo}, we assume the existence of a dummy 
value $\bot$ that is not a member of certain sets.
The following special notations concerning $\bot$ are used:
\begin{itemize}
\item
we write $\cA_\bot$, where $\cA$ is a set such that $\bot \notin \cA$, for 
$\cA \union \set{\bot}$;
\item
we write $\bot^n$, where $n \in \Natpos$, for the unique member of the 
set $\set{\bot}^n$.
\end{itemize}

In Section~\ref{sect-step-run-comp}, we consider multi-valued functions, 
i.e.\ functions from a set $\cA$ to the set of all subsets of a set 
$\cA'$.
The following notion concerning multi-valued functions is used:
\begin{udef}
Let $\cA$ and $\cA'$ be sets, and
let $f$ and $g$ be functions from $\cA$ to $\cP(\cA')$.
Then $f$ is \emph{smaller then} $g$ if $f(s) \subset g(s)$ for all 
$s \in \cA$.
\end{udef}

\section{The Informal Notion of a Concurrent Algorithm}
\label{sect-algo-informal}

In this paper, the notion of a classical algorithm concerns the original
notion of an algorithm, i.e.\ the notion that is intuitively 
characterized in standard works from the mathematical and computer 
science literature such as~\cite{Kle67a,Knu97a,Mal70a,Rog67a}, and
the notion of a concurrent algorithm concerns a generalization of the 
original notion of an algorithm.
Seeing the terminology used for proposed generalizations of the original 
notion of an algorithm, a classical algorithm could also be called a 
deterministic sequential non-interactive algorithm.

The characterizations of the informal notion of a classical algorithm 
referred to above indicate that a classical algorithm is considered to 
express a pattern of behaviour by which all instances of a computational 
problem can be solved.
This calls for a description of a classical computational problem that 
does not refer to the notion of a classical algorithm:
\begin{quote}
A classical computational problem is a problem where, given a value from a 
certain set of input values, a value from a certain set of output values 
that is in a certain relation to the given input value must be produced if 
it exists.
The input values are also called the instances of the problem and an 
output value that is in the certain relation to a given input value is 
also called a solution for the instance concerned.
\end{quote}

From the existing viewpoints on what a classical algorithm is, it 
follows that the following properties must be considered the most 
important ones of a classical algorithm: 
\begin{enumerate}
\item
a classical algorithm is a finite expression of a pattern of behaviour 
by which all instances of a classical computational problem can be 
solved;
\item
the pattern of behaviour expressed by a classical algorithm is made up 
of discrete steps, each of which consists of performing an elementary 
operation or inspecting an elementary condition unless it is the initial 
step or a final step;
\item
the pattern of behaviour expressed by a classical algorithm is such that 
there is one possible step immediately following a step that consists of 
performing an operation;
\item
the pattern of behaviour expressed by a classical algorithm is such that 
there is one possible step immediately following a step that consists of 
inspecting a condition for each outcome of the inspection;
\item
the pattern of behaviour expressed by a classical algorithm is such that 
the initial step consists of inputting an input value of the problem 
concerned;
\item
the pattern of behaviour expressed by a classical algorithm is such 
that, for each input value of the problem concerned for which a correct 
output value exists, a final step is reached after a finite number of 
steps and that final step consists of outputting a correct output value 
for that input value;
\item
the steps involved in the pattern of behaviour expressed by a classical 
algorithm are precisely and unambiguously defined and can be performed 
exactly in a finite amount of time.
\end{enumerate}

The notion of a classical algorithm covers algorithms that express a 
pattern of sequential behaviour. 
It does not cover algorithms in which parts can take place concurrently.
The notion of a concurrent algorithm covers algorithms that express a 
pattern of concurrent behaviour. 
A concurrent algorithm can be roughly characterized as consisting of a 
number of components that are classical algorithms except that:
\begin{itemize}
\item
there is only one component whose initial step consists of inputting an 
input value, there is only one component whose final steps consist of 
outputting an output value, and those components are the same;
\item
some elementary operations that can be performed by a component 
involve two values: a value available only to the component concerned 
and a value available to all components.
\end{itemize}

The above characterization has been derived from the viewpoints on 
concurrent computation that are expressed in computer science 
publications such as~\cite{CZ89a,FW78a,Gol82a,KRS90a}.
It reflects a rather operational view of what a concurrent algorithm is.
In a more abstract view of what a concurrent algorithm is, a 
concurrent algorithm expresses a collection of patterns of concurrent 
behaviour that are equivalent in some well-defined way.
We will come back to this at the end of Section~\ref{sect-proto-algo}.

\section{Concurrent Proto-Algorithms}
\label{sect-proto-algo}

In this section, the notion of a concurrent proto-algorithm is 
introduced.
The thought is that concurrent algorithms are equivalence classes of 
concurrent proto-algorithms under an appropriate equivalence relation.
An equivalence relation that is likely an appropriate one is introduced
in Section~\ref{sect-equivalence}.

We proceed with defining the three auxiliary notions, starting with the 
notion of an alphabet.
This notion concerns the symbols used to refer to the operations and 
conditions involved in the steps of which the pattern of behaviour 
expressed by a concurrent algorithm is made up.

\begin{udef}
An \emph{alphabet} $\Sigma$ is a quadruple $(F,\setf{F},\getf{F},P)$, 
where:
\begin{itemize}
\item
$F$ is a countable set of 
\emph{processing function symbols} of $\Sigma$;
\item
$\setf{F}$ is a countable set of
\emph{setting function symbols} of $\Sigma$;
\item
$\getf{F}$ is a countable set of 
\emph{getting function symbols} of $\Sigma$;
\item
$P$ is a countable set of 
\emph{predicate symbols} of $\Sigma$;
\item
$F$, $\setf{F}$, $\getf{F}$, and $P$ are disjoint sets and 
$\ini,\fin \in F$.
\end{itemize}
$\Sigma$ is called a \emph{classical} alphabet if $\setf{F} = \emptyset$
and $\getf{F} = \emptyset$.
\end{udef}
We write $\widehat{F}$ and $\procf{F}$, where $F$ is the set of 
processing function symbols of an alphabet, for the sets 
$F \diff \set{\fin}$ and $F \diff \set{\ini,\fin}$, respectively.

The function symbols and predicate symbols of an alphabet refer to the 
operations and conditions, respectively, involved in the steps of which 
the pattern of behaviour expressed by a concurrent algorithm is made 
up.
The function symbols $\ini$ and $\fin$ refer to inputting a first input 
value and outputting a last output value, respectively.

We are now ready to define the notions of a 
concurrent-$\Sigma$-algorithm component graph and a 
$\Sigma$-interpretation.
They concern the pattern of behaviour expressed by a concurrent
algorithm.

\begin{udef}
Let $\Sigma = (F,\setf{F},\getf{F},P)$ be an alphabet.
Then a \emph{concurrent-$\Sigma$-algo\-rithm component graph} $G$ is a 
rooted labeled directed graph $(V,E,\LBLv,\LBLe,l,r)$ such that
\begin{itemize}
\item
$\LBLv = F \union \setf{F} \union \getf{F} \union P$;
\item
$\LBLe = \set{0,1}$;
\item
for all $v \in V$:
\begin{itemize}
\item
$\indeg(v) = 0$ iff $v = r$;
\item
if $l(v) = \ini$, then $\indeg(v) = 0$;
\item
if $l(v) = \fin$, then $\outdeg(v) = 0$;
\item
if $l(v) \in \widehat{F} \union \setf{F} \union \getf{F}$, then
$\outdeg(v) = 1$ and, for the unique $v' \in V$ such that 
$(v,v') \in E$, $l((v,v'))$ is undefined;
\item
if $l(v)  \in P$, then
$\outdeg(v) = 2$ and, for the unique $v',v'' \in V$ such that 
$v' \neq v''$ and $(v,v'),(v,v'') \in E$, both $l((v,v'))$ and 
$l((v,v''))$ are defined and $l((v,v')) \neq l((v,v''))$;
\end{itemize}
\item
$l(r) = \ini$ iff there exists a $v \in V$ such that $\l(v) = \fin$;
\item
for all cycles $v_1\, \ldots\, v_{n+1}$ in $G$, 
there exists a $v \in \set{v_1,\ldots,v_n}$ such that 
$l(v) \in \procf{F} \union \setf{F} \union \getf{F}$.
\end{itemize}
$G$ is called a concurrent-$\Sigma$-algo\-rithm \emph{main} component 
graph if $l(r) = \ini$.
\end{udef}

In the above definition, the condition on cycles in a 
concurrent-$\Sigma$-algorithm component graph excludes infinitely many 
consecutive steps, each of which consists of inspecting a condition.

In the above definition, the conditions regarding the vertices and edges
of a concurrent-$\Sigma$-algorithm component graph correspond to the 
essential properties of the components of a concurrent algorithm 
described in Section~\ref{sect-algo-informal} that concern their 
structure.
Adding an interpretation of the symbols of the alphabet $\Sigma$ to a
concurrent-$\Sigma$-algorithm component graph yields something that has 
all of the essential properties of the components of a concurrent 
algorithm described in Section~\ref{sect-algo-informal}.

\begin{udef}
Let $\Sigma = (F,\setf{F},\getf{F},P)$ be an alphabet.
Then a \emph{$\Sigma$-interpretation} $\cI$ is a quadruple 
$(D,\Din,\Dout,I)$, 
where:
\begin{itemize}
\item
$D$ is a set, called the \emph{algorithm domain} of $\cI$;
\item
$\Din$ is a finitely generated set, called the \emph{input domain} of 
$\cI$;
\item
$\Dout$ is a finitely generated set, called the \emph{output domain} of 
$\cI$;
\item
$I$ is a total function from 
$F \union \setf{F} \union \getf{F} \union P$ to the set of all total 
computable functions from $\Din$ to $D$, $D$ to $\Dout$, $D$ to $D$, 
$D \x D$ to $D$ or $D$ to $\set{0,1}$ such that:
\begin{itemize}
\item
$I(\ini)$ is a function from $\Din$ to $D$;
\item
$I(\fin)$ is a function from $D$ to $\Dout$;
\item
for all $f \in \procf{F}$, $I(f)$ is a function from $D$ to $D$;
\item
for all $f \in \setf{F} \union \getf{F}$, $I(f)$ is a function from 
$D \x D$ to $D$;
\item
for all $p \in P$, $I(p)$ is a function from $D$ to $\set{0,1}$; 
\end{itemize}
\item
there does not exist a $D' \subset D$ such that:
\begin{itemize}
\item
for all $\din \in \Din$, $I(\ini)(\din) \in D'$;
\item
for all $f \in \procf{F}$, for all $d \in D'$, $I(f)(d) \in D'$;
\item
for all $f \in \setf{F} \union \getf{F}$, for all $d,d' \in D'$, 
$I(f)(d,d') \in D'$;
\end{itemize}
\item
for all $d,d' \in D$:
\begin{itemize}
\item
for all $f \in \setf{F}$,
$f(d,f(d,d')) = f(d,d')$ and
there exists a $f' \in \getf{F}$ such that $f'(d,f(d,d')) = d$;
\item
for all $f \in \getf{F}$,
$f(f(d,d'),d') = f(d,d')$ and
there exists a $f' \in \setf{F}$ such that $f'(f(d,d'),d') = d'$.
\end{itemize}
\end{itemize}
\end{udef}
The finite generation condition on $\Din$ and $\Dout$ and the minimality 
condition on $D$ in the above definition is considered desirable. 
They guarantee that all elements of $\Din$, $\Dout$, and $D$ have a 
finite representation, which is generally expected of the values 
involved in the steps of an algorithm. 
However, these conditions have not yet been shown to be essential in 
establishing results.
The condition on $\setf{F}$ and $\getf{F}$ in the above definition is 
considered characteristic of the nature of the setting functions and 
matching getting functions, but has not yet been shown to be essential 
in establishing results.

The pattern of behavior expressed by a concurrent algorithm can be 
fully represented by the combination of an alphabet $\Sigma$, a number 
of concurrent-$\Sigma$-algorithm component graphs $G_1,\ldots,G_n$, and 
a $\Sigma$-interpretation $\cI$.
This brings us to defining the notion of a concurrent proto-algorithm. 

\begin{udef}
A \emph{concurrent proto-algorithm} $A$ is a triple $(\Sigma,\vecG,\cI)$, 
where:
\begin{itemize}
\item
$\Sigma$ is an alphabet, called the \emph{alphabet} of $A$;
\item
$\vecG$ is a tuple of concurrent-$\Sigma$-algorithm component graphs, 
called the \emph{algorithm component graphs} of $A$, of which 
exactly one is a concurrent-$\Sigma$-algorithm main component graph;
\item
$\cI$ is a $\Sigma$-interpretation, 
called the \emph{interpretation} of $A$.
\end{itemize}
\end{udef}

In~\cite{Mid24a}, a definition of the notion of a classical 
proto-algorithm is given.%
\footnote
{Classical proto-algorithms are simply called proto-algorithms
 in~\cite{Mid24a}.}
A corollary of that definition and the definition of a concurrent 
proto-algorithm given above is that a classical proto-algorithm is a 
special case of a concurrent proto-algorithm.
\begin{corollary}
\label{corollary-classical}
A concurrent proto-algorithm $(\Sigma,\vecG,\cI)$ is a classical 
proto-algo\-rithm (as defined in~\cite{Mid24a}) 
if $\Sigma$ is a classical alphabet and $\vecG$ is a tuple of one 
concurrent-$\Sigma$-algorithm component graph.
\end{corollary}
Many definitions and results concerning concurrent proto-algorithms 
from the current paper are generalizations of definitions and results 
from~\cite{Mid24a}. 

Henceforth, we assume a dummy value $\bot$ such that, 
for all concurrent proto-algorithms $A = (\Sigma,\vecG,\cI)$, where 
$\Sigma = (F,\setf{F},\getf{F},P)$, $\vecG = (G_1,\ldots,G_n)$ with
$G_i = (V_i,E_i,\LBLv_i,\LBLe_i,l_i,r_i)$ ($1 \leq i \leq n$), and 
$\cI = (D,\Din,\Dout,I)$, $\bot \notin V_i$ for each 
$i \in \set{1,\ldots,n}$, $\bot \notin D$, $\bot \notin \Din$, 
$\bot \notin \Dout$, 
and $\bot \notin \set{0,\ldots,n}$.

The intuition is that a concurrent proto-algorithm is something that 
goes from one state to another.
\begin{udef} 
Let $A = (\Sigma,\vecG,\cI)$ be a concurrent proto-algorithm, where 
$\Sigma = (F,\setf{F},\getf{F},P)$, $\vecG = (G_1,\ldots,G_n)$ with
$G_i = (V_i,E_i,\LBLv_i,\LBLe_i,l_i,r_i)$ ($1 \leq i \leq n$), and 
$\cI = (D,\Din,\Dout,I)$.
Then a \emph{state} of $A$ is a triple $(\din,c,\dout)$, where
$\din \in \Din_\bot$, 
$c \in
 ({V_1}_\bot \x \ldots \x {V_n}_\bot) \x {D_\bot}^n \x {D_\bot} \x
 \set{1,\ldots,n}_\bot$, and
$\dout \in \Dout_\bot$,
such that:
\begin{itemize}
\item
if $c = (\bot^n,\bot^n,\bot,\bot)$, then 
$\din = \bot$ \,\,iff\,\, $\dout \neq \bot$;
\item
if $c \neq (\bot^n,\bot^n,\bot,\bot)$, then 
$\din = \bot$ and $\dout = \bot$.
\end{itemize}
A state $(\din,c,\dout)$ of $A$ is called an \emph{initial state} of $A$ 
if $\din \neq \bot$,
a state $(\din,c,\dout)$ of $A$ is called a \emph{final state} of $A$ if
$\dout \neq \bot$, and 
a state $(\din,c,\dout)$ of $A$ is called an \emph{internal state} of $A$ 
if $\din  = \bot$ and $\dout  = \bot$. 
\end{udef}
Henceforth, we write $\cS_A$, where $A$ is a concurrent 
proto-algorithm, for the set of all states of $A$. 
We also write $\Sini_A$ for the set of all initial states of $A$, 
$\Sfin_A$ for the set of all final states of $A$, and $\Sint_A$ for the 
set of all internal states of $A$.
Moreover, we write $\botc$ for the tuple $(\bot^n,\bot^n,\bot,\bot)$.

An internal state $((v_1,\ldots,v_n),(d_1,\ldots,d_n),d,j)$ of a 
proto-algorithm $A$ can be largely explained from the point of view of 
state machines:
\begin{itemize}
\item
$v_i$ is the control state of the $i$th component of $A$
($1 \leq i \leq n$);
\item
$d_i$ is the private data state of the $i$th component of $A$
($1 \leq i \leq n$);
\item
$d$ is the shared data state of the components of $A$.
\end{itemize}
Moreover, $j$ indicates that this state is one in which a step to a next 
state is made by the $j$th component of $A$.

Suppose that $A = (\Sigma,\vecG,\cI)$ is a concurrent proto-algorithm, 
where $\Sigma = (F,\setf{F},\getf{F},P)$, $\vecG = (G_1,\ldots,G_n)$ with
$G_i = (V_i,E_i,\LBLv_i,\LBLe_i,l_i,r_i)$ ($1 \leq i \leq n$),   
$\cI = (D,\Din,\Dout,I)$, and $\vecr = (r_1,\ldots,r_n)$.
$A$ goes from one state to the next state by making a step, it starts in 
an initial state, and if it does not keep making steps forever, it stops 
in a final state.
The following is an informal explanation of how the state that $A$ is in 
determines what the possible steps to a next state consists of and to 
what next states they lead:
\begin{itemize}
\newpage
\sloppy
\item
if $A$ is in initial state $(\din,\botc,\bot)$ and 
$i \in \set{1,\ldots,n}$ is such that $l(\vecr(i)) = \ini$, then, 
for each $j \in \set{1,\ldots,n}$ with $\outdeg(\vecr(j)) > 0$, a step 
to a next state is possible that consists of applying function $I(\ini)$ 
to $\din$ and leads to the unique internal state 
$(\bot,(\updtup{\vecr}{i}{v'},\updtup{\bot^n}{i}{d'},\bot,j),\bot)$ 
such that $(\vecr(i),v') \in E_i$ and $I(\ini)(\din) = d'$;
\item
if $A$ is in internal state $(\bot,(\vecv,\vecd,d,i),\bot)$ and
$l(\vecv(i)) \in \procf{F}$, then, 
for each $j \in \set{1,\ldots,n}$ with $\outdeg(\vecv(j)) > 0$, a step to a next state 
is possible that consists of applying function $I(l(\vecv(i)))$ to 
$\vecd(i)$ and leads to the unique internal state 
$(\bot,(\updtup{\vecv}{i}{v'},\linebreak[2]
        \updtup{\vecd}{i}{d'},d,j),\bot)$ 
such that $(\vecv(i),v') \in E_i$, and $I(l(\vecv(i)))(\vecd(i)) = d'$;
\item
if $A$ is in internal state $(\bot,(\vecv,\vecd,d,i),\bot)$ and 
$l(\vecv(i)) \in \setf{F}$, then, 
for each $j \in \set{1,\ldots,n}$ with $\outdeg(\vecv(j)) > 0$, a step to a next state is possible 
that consists of applying function $I(l(\vecv(i)))$ to $(\vecd(i),d)$ 
and leads to the unique internal state 
$(\bot,(\updtup{\vecv}{i}{v'},\linebreak[2]\vecd,d',j),\bot)$ 
such that $(\vecv(i),v') \in E_i$, and 
$I(l(\vecv(i)))(\vecd(i),d) = d'$;
\item
if $A$ is in internal state $(\bot,(\vecv,\vecd,d,i),\bot)$ and 
$l(\vecv(i)) \in \getf{F}$, then,
for each $j \in \set{1,\ldots,n}$ with $\outdeg(\vecv(j)) > 0$, a step to a next state is possible 
that consists of applying function $I(l(\vecv(i)))$ to $(\vecd(i),d)$ 
and leads to the unique internal state 
$(\bot,(\updtup{\vecv}{i}{v'},\linebreak[2]
        \updtup{\vecd}{i}{d'},d,j),\bot)$ 
such that $(\vecv(i),v') \in E_i$, and 
$I(l(\vecv(i)))(\vecd(i),d) = d'$;
\item
if $A$ is in internal state $(\bot,(\vecv,\vecd,d,i),\bot)$ and 
$l(\vecv(i)) \in P$, then a step to a next state is 
possible that consists of applying function $I(l(\vecv(i)))$ to 
$\vecd(i)$ and leads to the unique internal state 
$(\bot,(\updtup{\vecv}{i}{v'},\vecd,d,i),\bot)$ 
such that $(\vecv(i),v') \in E_i$, and 
$I(l(\vecv(i)))(\vecd(i)) = l((\vecv(i),v'))$;
\item
if $A$ is in internal state $(\bot,(\vecv,\vecd,d,i),\bot)$, 
$l(\vecv(i)) = \fin$, and for all $j \in \set{1,\ldots,n}$, $\outdeg(\vecv(j)) = 0$, then a step to a next state is 
possible that consists of applying function $I(\fin)$ to $\vecd(i)$ and 
leads to the unique final state $(\bot,\botc,\dout)$ 
such that $I(\fin)(\vecd(i)) = \dout$.
\end{itemize}
This informal explanation of how the state that $A$ is in determines 
what the possible next states are, is formalized by the algorithmic step 
function $\astep_A$ defined in Section~\ref{sect-step-run-comp}. 

Because concurrent proto-algorithms are considered too concrete to be 
called concurrent algorithms, the term concurrent proto-algorithm has 
been chosen instead of the term concurrent algorithm.
For example, from a mathematical point of view, it is natural to 
consider the behavioral patterns expressed by isomorphic concurrent 
proto-algorithms to be the same.

Although it is intuitive clear what isomorphism of concurrent 
proto-algorithms is, its precise definition is long and tedious.
\begin{udef}
\sloppy
Let $A = (\Sigma,\vecG,\cI)$ and $A' = (\Sigma',\vecG',\cI')$ be 
concurrent proto-algorithms, where
$\Sigma = (F,\setf{F},\getf{F},P)$, 
$\Sigma' = (F',\setf{F}',\getf{F}',P')$, 
$\vecG = (G_1,\ldots,G_n)$ with
$G_i = (V_i,E_i,\LBLv_i,\LBLe_i,l_i,r_i)$ ($1 \leq i \leq n$), 
$\vecG' = (G'_1,\ldots,G'_{n'})$ with
$G'_j = (V'_j,E'_j,\LBLv'_j,\LBLe'_j,l'_j,r'_j)$ 
($1 \leq j \leq n'$), 
$\cI = (D,\Din,\Dout,I)$, and $\cI' = (D',\Din',\Dout',I')$.
Then $A$ and $A'$ are \emph{isomorphic}, written $A \iso A'$, 
if there exist bijections 
$\funct{\bijX}{\set{1,\ldots,n}}{\set{1,\ldots,n}}$, 
$\funct{\bijF}{F}{F'}$, $\funct{\bijS}{\setf{F}}{\setf{F}'}$, 
$\funct{\bijG}{\getf{F}}{\getf{F}'}$, $\funct{\bijP}{P}{P'}$, 
$\funct{\bijV_i}{V_i}{V'_{\bijX(i)}}$ for $i \in \set{1,\ldots,n}$, 
$\funct{\bijD}{D}{D'}$, 
$\funct{\bijI}{\Din}{\Din'}$, $\funct{\bijO}{\Dout}{\Dout'}$, and 
$\funct{\bijB}{\set{0,1}}{\set{0,1}}$ such that:
\begin{itemize}
\item
$n = n'$;
\item
for all $i \in \set{1,\ldots,n}$:
\begin{itemize}  
\item
for all $v,v' \in V_i$,  
$(v,v') \in E_i$ iff $(\bijV_i(v),\bijV_i(v')) \in E'_{\bijX(i)}$;
\item
for all $v \in V_i$ with $l_i(v) \in F$,\, 
$\bijF(l_i(v)) = l'_{\bijX(i)}(\bijV_i(v))$;
\item
for all $v \in V_i$ with $l_i(v) \in \setf{F}$, 
$\bijS(l_i(v)) = l'_{\bijX(i)}(\bijV_i(v))$;
\item
for all $v \in V_i$ with $l_i(v) \in \getf{F}$, 
$\bijG(l_i(v)) = l'_{\bijX(i)}(\bijV_i(v))$;
\item
for all $v \in V_i$ with $l_i(v) \in P$,\, 
$\bijP(l_i(v)) = l'_{\bijX(i)}(\bijV_i(v))$; 
\item
for all $(v,v') \in E_i$ with $l_i((v,v'))$ defined, \\ \hspace*{1.25em}
$\bijB(l_i((v,v'))) = l'_{\bijX(i)}((\bijV_i(v),\bijV_i(v')))$;
\end{itemize}
\item
$\bijF(\ini) = \ini$ and $\bijF(\fin) = \fin$;
\item
for all $\din \in \Din$, $\bijD(I(\ini)(\din)) = I'(\ini)(\bijI(\din))$;
\item
for all $d \in D$, $\bijO(I(\fin)(d)) = I'(\fin)(\bijD(d))$;
\item
for all $d \in D$ and $f \in \procf{F}$, 
$\bijD(I(f)(d)) = I'(\bijF(f))(\bijD(d))$;
\item
for all $d,d' \in D$ and $f \in \setf{F}$, 
$\bijD(I(f)(d,d')) = I'(\bijS(f))(\bijD(d),\bijD(d'))$;
\item
for all $d,d' \in D$ and $f \in \getf{F}$, 
$\bijD(I(f)(d,d')) = I'(\bijG(f))(\bijD(d),\bijD(d'))$;
\item
for all $d \in D$ and $p \in P$, 
$\bijB(I(p)(d)) = I'(\bijP(p))(\bijD(d))$.
\end{itemize}
\end{udef}

Concurrent proto-algorithms may also be considered too concrete in a 
way not covered by isomorphism of concurrent proto-algorithms.
This issue is addressed in Section~\ref{sect-equivalence} and leads 
there to the introduction of two other equivalence relations.

\sloppy
A concurrent proto-algorithm could also be defined as a quadruple 
$(D,\Din,\Dout,\overline{\vecG})$ where $\overline{\vecG}$ is a tuple of
graphs that differ from concurrent-$\Sigma$-algorithm component graphs 
in that their vertex labels are computable functions from $\Din$ to $D$, 
$D$ to $\Dout$, $D$ to $D$, $D \x D$ to $D$ or $D$ to $\set{0,1}$ 
instead of function and predicate symbols from $\Sigma$.
I consider the definition of a concurrent proto-algorithm given earlier 
more insightful because it isolates as much as possible the operations 
to be performed and the conditions to be inspected from the structure of
a concurrent proto-algorithm.

\section{On Steps, Runs and What is Computed}
\label{sect-step-run-comp}

In Section~\ref{sect-proto-algo}, the intuition was given that a 
concurrent proto-algorithm $A$ is something that goes through states.
It was informally explained how the state that it is in determines what 
the possible steps to a next state consists of and to what next states 
they lead.
The algorithmic step function $\astep_A$ that is defined below 
formalizes this.
The computational step function $\cstep_A$ that is also defined below is 
like the algorithmic step function $\astep_A$, but conceals the steps 
that consist of inspecting conditions.

\begin{udef}
Let $A = (\Sigma,\vecG,\cI)$ be a concurrent proto-algorithm, 
where $\Sigma = (F,\setf{F},\getf{F},P)$, $\vecG = (G_1,\ldots,G_n)$ with
$G_i = (V_i,E_i,\LBLv_i,\LBLe_i,l_i,r_i)$ ($1 \leq i \leq n$), and
$\cI = (D,\Din,\Dout,I)$, and let $\vecr = (r_1,\ldots,r_n)$.
Then the \emph{algorithmic step function $\astep_A$ induced by $A$} is 
the smallest total function from $\cS_A$ to $\cP(\cS_A) \diff \emptyset$ 
such that for all $d,d' \in D$, $\din \in \Din$, $\dout \in \Dout$, 
$v_1' \in V_1$, \ldots, $v_n' \in V_n$, $\vecd \in D^n$, 
$\vecv \in V_1 \x \ldots \x V_n$, and $i,j \in \set{1,\ldots,n}$:%
\footnote{In this definition, we write $S \ni s$ instead of $s \in S$.}
\begin{center}
\renewcommand{\arraystretch}{1}
\begin{tabular}{@{}r@{}l@{\,}l@{}} 
$\astep_A((\din,\botc,\bot))$ & 
${} \ni 
 (\bot,(\updtup{\vecr}{i}{v'_i},\updtup{\bot^n}{i}{d'},\bot,j),\bot)$ \\
\multicolumn{3}{l}{\hspace*{3.6em}
if $l(\vecr(i)) = \ini$, $(\vecr(i),v'_i) \in E_i$,
      $I(\ini)(\din) = d'$, and $\outdeg(\vecr(j)) > 0$,}
\\
$\astep_A((\bot,(\vecv,\vecd,d,i),\bot))$ & 
${} \ni
 (\bot,(\updtup{\vecv}{i}{v'_i},\updtup{\vecd}{i}{d'},d,j),\bot)$ \\ 
\multicolumn{3}{l}{\hspace*{3.6em}
if $l(\vecv(i)) \in \procf{F}$, $(\vecv(i),v'_i) \in E_i$,  
$I(l(\vecv(i)))(\vecd(i)) = d'$, and}
\\
\multicolumn{3}{l}{\hspace*{3.6em} \phantom{if}
$\outdeg(\vecv(j)) > 0$,}
\\
$\astep_A((\bot,(\vecv,\vecd,d,i),\bot))$ & 
${} \ni (\bot,(\updtup{\vecv}{i}{v'_i},\vecd,d',j),\bot)$ \\ 
\multicolumn{3}{l}{\hspace*{3.6em}
if $l(\vecv(i)) \in \setf{F}$, $(\vecv(i),v'_i) \in E_i$,  
$I(l(\vecv(i)))(\vecd(i),d) = d'$, and}
\\
\multicolumn{3}{l}{\hspace*{3.6em} \phantom{if}
$\outdeg(\vecv(j)) > 0$,}
\\
$\astep_A((\bot,(\vecv,\vecd,d,i),\bot))$ & 
${} \ni (\bot,(\updtup{\vecv}{i}{v'_i},\updtup{\vecd}{i}{d'},d,j),\bot)$ \\ 
\multicolumn{3}{l}{\hspace*{3.6em}
if $l(\vecv(i)) \in \getf{F}$, $(\vecv(i),v'_i) \in E_i$,  
$I(l(\vecv(i)))(\vecd(i),d) = d'$, and}
\\
\multicolumn{3}{l}{\hspace*{3.6em} \phantom{if}
$\outdeg(\vecv(j)) > 0$,}
\\
$\astep_A((\bot,(\vecv,\vecd,d,i),\bot))$ & 
${} \ni
 (\bot,(\updtup{\vecv}{i}{v'_i},\vecd,d,i),\bot)$ \\
\multicolumn{3}{l}{\hspace*{3.6em}
if $l(\vecv(i)) \in P$, $(\vecv(i),v'_i) \in E_i$, and 
$I(l(\vecv(i)))(\vecd(i)) = l((\vecv(i),v'_i))$,}
\\
$\astep_A((\bot,(\vecv,\vecd,d,i),\bot))$ & 
${} \ni (\bot,\botc,\dout)$ \\ 
\multicolumn{3}{l}{\hspace*{3.6em}
if $l(\vecv(i)) = \fin$, $I(\fin)(\vecd(i)) = \dout$, and}
\\
\multicolumn{3}{l}{\hspace*{3.6em} \phantom{if}
for all $k \in \set{1,\ldots,n}$, $\outdeg(\vecv(k)) = 0$,}
\\
$\astep_A((\bot,\botc,\dout))$ & 
${} \ni (\bot,\botc,\dout)$
\end{tabular}
\end{center}
and the \emph{computational step function $\cstep_A$ induced by $A$} is 
the smallest total function from $\cS_A$ to $\cP(\cS_A) \diff \emptyset$ 
such that, for all $d,d' \in D$, $\din \in \Din$, $\dout \in \Dout$, 
\linebreak[2] $v_1' \in V_1$, \ldots, $v_n' \in V_n$, $\vecd \in D^n$, 
$\vecv \in V_1 \x \ldots \x V_n$, $i,j \in \set{1,\ldots,n}$, and 
$s \in \cS_A$:
\begin{center}
\renewcommand{\arraystretch}{1}
\begin{tabular}{@{}r@{}l@{\,}l@{}} 
$\cstep_A((\din,\botc,\bot))$ & 
${} \ni 
 (\bot,(\updtup{\vecr}{i}{v'_i},\updtup{\bot^n}{i}{d'},\bot,j),\bot)$ \\
\multicolumn{3}{l}{\hspace*{3.6em}
if $l(\vecr(i)) = \ini$, $(\vecr(i),v'_i) \in E_i$, 
      $I(\ini)(\din) = d'$, and $\outdeg(\vecr(j)) > 0$,}
\\
$\cstep_A((\bot,(\vecv,\vecd,d,i),\bot))$ & 
${} \ni
 (\bot,(\updtup{\vecv}{i}{v'_i},\updtup{\vecd}{i}{d'},d,j),\bot)$ \\ 
\multicolumn{3}{l}{\hspace*{3.6em}
if $l(\vecv(i)) \in \procf{F}$, $(\vecv(i),v'_i) \in E_i$,  
$I(l(\vecv(i)))(\vecd(i)) = d'$, and}
\\
\multicolumn{3}{l}{\hspace*{3.6em} \phantom{if}
$\outdeg(\vecv(j)) > 0$,}
\\
$\cstep_A((\bot,(\vecv,\vecd,d,i),\bot))$ & 
${} \ni (\bot,(\updtup{\vecv}{i}{v'_i},\vecd,d',j),\bot)$ \\ 
\multicolumn{3}{l}{\hspace*{3.6em}
if $l(\vecv(i)) \in \setf{F}$, $(\vecv(i),v'_i) \in E_i$,  
$I(l(\vecv(i)))(\vecd(i),d) = d'$, and}
\\
\multicolumn{3}{l}{\hspace*{3.6em} \phantom{if}
$\outdeg(\vecv(j)) > 0$,}
\\
$\cstep_A((\bot,(\vecv,\vecd,d,i),\bot))$ & 
${} \ni (\bot,(\updtup{\vecv}{i}{v'_i},\updtup{\vecd}{i}{d'},d,j),\bot)$ \\ 
\multicolumn{3}{l}{\hspace*{3.6em}
if $l(\vecv(i)) \in \getf{F}$, $(\vecv(i),v'_i) \in E_i$,  
$I(l(\vecv(i)))(\vecd(i),d) = d'$, and}
\\
\multicolumn{3}{l}{\hspace*{3.6em} \phantom{if}
$\outdeg(\vecv(j)) > 0$,}
\\
$\cstep_A((\bot,(\vecv,\vecd,d,i),\bot))$ & ${} \ni s$ \\
\multicolumn{3}{l}{\hspace*{3.6em}
if $l(\vecv(i)) \in P$, $(\vecv(i),v'_i) \in E_i$,  
$I(l(\vecv(i)))(\vecd(i)) = l((\vecv(i),v'_i))$, and} \\
\multicolumn{3}{l}{\hspace*{3.6em} \phantom{if}
$\cstep_A((\bot,(\updtup{\vecv}{i}{v'_i},\vecd,d,i),\bot))$
${} \ni s$,}
\\
$\cstep_A((\bot,(\vecv,\vecd,d,i),\bot))$ & 
${} \ni (\bot,\botc,\dout)$ \\ 
\multicolumn{3}{l}{\hspace*{3.6em}
if $l(\vecv(i)) = \fin$, $I(\fin)(\vecd(i)) = \dout$, and}
\\
\multicolumn{3}{l}{\hspace*{3.6em} \phantom{if}
for all $k \in \set{1,\ldots,n}$, $\outdeg(\vecv(k)) = 0$,}
\\
$\cstep_A((\bot,\botc,\dout))$ & 
${} \ni (\bot,\botc,\dout)$.
\end{tabular}
\end{center}
\end{udef}

Below, we define what the possible runs of a concurrent proto-algorithm 
on an input value is and what is computed by a concurrent 
proto-algorithm.
To this end, we first give some auxiliary definitions.

In the coming definitions, we write
$\cA^\infty$ for the set of all non-empty sequences over the set $\cA$ 
that are finite or countably infinite,
$\emptyseq$ for the empty sequence,
$\seq{a}$ for the sequence having $a$ as sole element, 
$\alpha \conc \alpha'$ for the concatenation of the sequences $\alpha$ 
and $\alpha'$, and 
$|\alpha|$ for the length of the sequence $\alpha$.

\begin{udef}
Let $A = (\Sigma,\vecG,\cI)$ be a concurrent proto-algorithm.
Then the \emph{algorithmic semi-run set function $\asruns_A$ induced by 
$A$} and the \emph{computational semi-run set function $\csruns_A$ 
induced by $A$} are the total functions from $\cS_A$ to 
$\cP({\cS_A}^\infty)$ such that for all $s \in \cS_A$:
\begin{center}
\renewcommand{\arraystretch}{1}
\hspace*{1.5em}
\begin{tabular}{@{}l@{}l@{\;}l@{}} 
$\asruns_A(s)$ & ${} = 
 \set{\seq{s} \conc \sigma \where
      \Lexists{s' \in \astep_A(s)}{\sigma \in \asruns_A(s')}}$ &
if $s \notin \Sfin_A$,
\\
$\asruns_A(s)$ & ${} = \set{\seq{s}}$ & 
if $s \in \Sfin_A$.
\end{tabular}
\end{center} 
and
\begin{center}
\renewcommand{\arraystretch}{1}
\hspace*{1.5em}
\begin{tabular}{@{}l@{}l@{\;}l@{}} 
$\csruns_A(s)$ & ${} = 
 \set{\seq{s} \conc \sigma \where
      \Lexists{s' \in \cstep_A(s)}{\sigma \in \csruns_A(s')}}$ &
if $s \notin \Sfin_A$,
\\
$\csruns_A(s)$ & ${} = \set{\seq{s}}$ & 
if $s \in \Sfin_A$.
\end{tabular}
\end{center} 
Moreover, the \emph{output value extraction function 
$\outputx_A$ for $A$} is the total function from ${\cS_A}^\infty$ to 
$\Dout_\bot$ such that for all $(\din,c,\dout) \in \cS_A$ and 
$\sigma \in {\cS_A}^\infty$: 
\begin{center}
\renewcommand{\arraystretch}{1}
\hspace*{1.5em}
\begin{tabular}{@{}l@{}l@{\;}l@{}} 
$\outputx_A(\seq{(\din,c,\dout)} \conc \sigma)$ & 
${} = \outputx_A(\sigma)$ & if $\dout = \bot$,
\\
$\outputx_A(\seq{(\din,c,\dout)} \conc \sigma)$ & 
${} = \dout$ &              if $\dout \neq \bot$, 
\\
$\outputx_A(\emptyseq)$ & ${} = \bot$.
\end{tabular}
\end{center} 
A $\sigma \in {\cS_A}^\infty$ is called an \emph{algorithmic semi-run} 
if there exists an $s \in \cS_A$ such that $\sigma \in \asruns_A(s)$. 
\end{udef}
Not every $\sigma \in {\cS_A}^\infty$ is an algorithmic semi-run.
Take, for example, a sequence $\sigma' \conc \seq{(\bot,c,\bot)}$, 
where $\sigma' \in {\cS_A}^\infty$ and $(\bot,c,\bot) \in \cS_A$.
Then $\sigma' \conc \seq{(\bot,c,\bot)} \in {\cS_A}^\infty$, but
$\sigma' \conc \seq{(\bot,c,\bot)}$ is not an algorithmic semi-run.
For each algorithmic semi-run $\sigma$, $\outputx_A(\sigma) \neq \bot$.

The following definition concerns what the possible runs of a concurrent 
proto-algorithm on an input value is.
As with steps, a distinction is made between algorithmic runs and 
computational runs.
\begin{udef}
Let $A = (\Sigma,\vecG,\cI)$ be a concurrent proto-algorithm, where
$\cI = (D,\Din,\Dout,I)$, and
let $\din \in \Din$.
Then the \emph{algorithmic run set of $A$ on $\din$}, 
written $\arun_A(\din)$, is $\asruns_A((\din,\botc,\bot))$ and
the \emph{computational run set of $A$ on $\din$}, 
written $\crun_A(\din)$, is $\csruns_A((\din,\botc,\bot))$.
\end{udef}

When defining what is computed by a concurrent proto-algorithm, a 
distinction must be made between convergent runs and divergent runs.
\begin{udef}
Let $A = (\Sigma,\vecG,\cI)$ be a concurrent proto-algorithm, and 
let $\sigma \in {\cS_A}^\infty$. 
Then \emph{$\sigma$ is divergent} if there exists a suffix $\sigma'$ of 
$\sigma$ such that $\sigma' \in {\Sint_A}^\infty$, and 
\emph{$\sigma$ is convergent} if $\sigma$ is not divergent.
\end{udef}

The following definition concerns what is computed by a concurrent 
proto-algorithm.
\begin{udef}
Let $A = (\Sigma,\vecG,\cI)$ be a concurrent proto-algorithm, where 
$\cI = (D,\Din,\Dout,I)$. 
Then the \emph{relation $\widehat{A}$ computed by $A$} is the relation 
from $\Din$ to $\Dout$ such that 
for all $\din \in \Din$ and $\dout \in \Dout$:
\begin{center}
\renewcommand{\arraystretch}{1.2}
\begin{tabular}{@{}l@{}} 
$(\din,\dout) \in \widehat{A}$ \,\,iff\,\, 
there exists a convergent $\sigma \in \arun_A(\din)$ such that 
$\dout = \outputx_A(\sigma)$.
\end{tabular}
\end{center}
\end{udef}

If $A = (\Sigma,\vecG,\cI)$ is a concurrent proto-algorithm where 
$\vecG$ is a tuple of only one concurrent-$\Sigma$-algorithm component 
graph, then the relation $\widehat{A}$ computed by $A$ is functional, 
i.e.\ $\widehat{A}$ is (the graph of) a partial function.

\section{Algorithmic and Computational Equivalence}
\label{sect-equivalence}

If a concurrent proto-algorithm $A'$ can mimic a concurrent 
proto-algorithm $A$ step-by-step, then we say that $A$ is 
algorithmically simulated by $A'$.
If the steps that consist of inspecting conditions are ignored, then we 
say that $A$ is computationally simulated by $A'$.
Algorithmic and computational simulation can be formally defined using 
the step functions defined in Section~\ref{sect-step-run-comp}.

\begin{udef}
Let $A = (\Sigma,\vecG,\cI)$ and $A' = (\Sigma',\vecG',\cI')$ be 
concurrent proto-algorithms, where
$\Sigma = (F,\setf{F},\getf{F},P)$, 
$\Sigma' = (F',\setf{F}',\getf{F}',P')$, 
$\vecG = (G_1,\ldots,G_n)$ with
$G_i = (V_i,E_i,\LBLv_i,\LBLe_i,l_i,r_i)$ ($1 \leq i \leq n$), 
$\vecG' = (G'_1,\ldots,G'_{n'})$ with
$G'_j = (V'_j,E'_j,\LBLv'_j,\LBLe'_j,l'_j,r'_j)$ 
($1 \leq j \leq n'$), 
$\cI = (D,\Din,\Dout,I)$, and $\cI' = (D',\Din',\Dout',I')$.
Then an \emph{algorithmic simulation of $A$ by $A'$} is a set
$R \subseteq \cS_A \x \cS_{A'}$ such that: 
\begin{itemize}
\item
for all $s \in \cS_A$ and $s' \in \cS_{A'}$:
\begin{itemize}
\item
if $\sini \in \Sini_A$, then there exists an 
$\sini' \in \Sini_{A'}$ such that $(\sini,\sini') \in R$;
\item
if $\sfin' \in \Sfin_{A'}$, then there exists an  
$\sfin \in \Sfin_A$ such that $(\sfin,\sfin') \in R$;
\item
if $(s,s') \in R$ and $t \in \astep_A(s)$, 
then there exists a $t' \in \astep_{A'}(s')$ such that 
{$(t,t') \in R$};
\end{itemize}
\item
for all $(s,s') \in R$:
\begin{itemize}
\item
$s \in \Sini_A$ iff $s' = \Sini_{A'}$;
\item 
$s \in \Sfin_A$ iff $s' = \Sfin_{A'}$
\end{itemize}
\end{itemize}
and a \emph{computational simulation of $A$ by $A'$} is a set
$R \subseteq \cS_A \x \cS_{A'}$ such that: 
\begin{itemize}
\item
for all $s \in \cS_A$ and $s' \in \cS_{A'}$:
\begin{itemize}
\item
if $\sini \in \Sini_A$, then there exists an 
$\sini' \in \Sini_{A'}$ such that $(\sini,\sini') \in R$;
\item
if $\sfin' \in \Sfin_{A'}$, then there exists an  
$\sfin \in \Sfin_A$ such that $(\sfin,\sfin') \in R$;
\item
if $(s,s') \in R$ and $t \in \cstep_A(s)$, 
then there exists a $t' \in \cstep_{A'}(s')$ such that 
{$(t,t') \in R$};
\end{itemize}
\item
for all $(s,s') \in R$:
\begin{itemize}
\item
$s \in \Sini_A$ iff $s' = \Sini_{A'}$;
\item 
$s \in \Sfin_A$ iff $s' = \Sfin_{A'}$.
\end{itemize}
\end{itemize}
$A$ \emph{is algorithmically simulated by} $A'$, written $A \asim A'$,
if there exists an algorithmic simulation $R$ of $A$ by $A'$. 
\\ 
$A$ \emph{is computationally simulated by} $A'$, written $A \csim A'$, 
if there exists a computational simulation $R$ of $A$ by $A'$.
\pagebreak[2]
\\
$A$ \emph{is algorithmically equivalent to} $A'$, written $A \aeqv A'$, 
if there exist an algorithmic simulation $R$ of $A$ by $A'$ and an 
algorithmic simulation $R'$ of $A'$ by $A$ such that $R' = R^{-1}$.
\\
$A$ \emph{is computationally equivalent to} $A'$, written $A \ceqv A'$, 
if there exist an computational simulation $R$ of $A$ by $A'$ and an 
computational simulation $R'$ of $A'$ by $A$ such that $R' = R^{-1}$.
\end{udef}

The conditions imposed on an algorithmic or computational simulation $R$ 
of a concurrent proto-algorithm $A$ by a concurrent proto-algorithm 
$A'$ include, in addition to the usual transfer conditions, also 
conditions that guarantee that a state of $A$ is only related by $R$ to 
a state of $A'$ of the same kind (initial, final or internal).

There may be states of a concurrent proto-algorithm in which there is a 
choice from multiple possible steps to a next state.
The condition $R' = R^{-1}$ imposed on simulations $R$ and $R'$ 
witnessing (algorithmic or computational) equivalence of concurrent 
proto-algorithms $A$ and $A'$ guarantees that $A$ and $A'$ have the same 
choice structure.

The next lemma will be used in the proof of the theorem that follows it.
In this lemma and the proof of the theorem that follows it, we write 
$\alpha[n]$ for the $n$th element of the sequence $\alpha$ if there 
exists a prefix of $\alpha$ with length $n$ and otherwise the last 
element of $\alpha$.
\begin{lemma}
\label{lemma-simulation}
Let $A = (\Sigma,\vecG,\cI)$ and $A' = (\Sigma',\vecG',\cI')$ be 
concurrent proto-algorithms, where
$\Sigma = (F,\setf{F},\getf{F},P)$, 
$\Sigma' = (F',\setf{F}',\getf{F}',P')$, 
$\vecG = (G_1,\ldots,G_n)$ with
$G_i = (V_i,E_i,\LBLv_i,\LBLe_i,l_i,r_i)$ ($1 \leq i \leq n$), 
$\vecG' = (G'_1,\ldots,G'_{n'})$ with
$G'_j = (V'_j,E'_j,\LBLv'_j,\LBLe'_j,l'_j,r'_j)$ 
($1 \leq j \leq n'$), 
$\cI = (D,\Din,\Dout,I)$, and $\cI' = (D',\Din',\Dout',I')$, and
let $R \subseteq \cS_A \x \cS_{A'}$ and 
$\funct{\fncI}{\Din}{\Din'}$ be such that, for all $\din \in \Din$, 
$((\din,\botc,\bot),(\fncI(\din),\botc,\bot)) \in R$.
Then $R$ is an algorithmic simulation of $A$ by $A'$ only if,
for all $\din \in \Din$, for all $\sigma \in \arun_A(\din)$,
there exists a $\sigma' \in \arun_{A'}(\fncI(\din))$ such that,
for all $n \in \Natpos$, $(\sigma[n],\sigma'[n]) \in R$. 
\end{lemma}
\begin{proof}
Let $R$ be an algorithmic simulation of $A$ by $A'$ and 
$\funct{\fncI}{\Din}{\Din'}$ be such that, for all $\din \in \Din$, 
$((\din,\botc,\bot),(\fncI(\din),\botc,\bot)) \in R$, and
let $\sigma \in \arun_A(\din)$.
Then we can easily construct a $\sigma' \in \arun_{A'}(\fncI(\din))$ 
such that, for all $n \in \Natpos$, $(\sigma[n],\sigma'[n]) \in R$,
using the transfer conditions imposed on algorithmic simulations. 
\qed
\end{proof}

The following theorem tells us that, 
if a concurrent proto-algorithm $A$ is algorithmically simulated by a 
concurrent proto-algorithm $A'$, then 
(a)~the relation computed by $A'$ models the relation computed by $A$ 
(in the sense of e.g.~\cite{Jon90a}) and
(b)~for each convergent algorithmic run of $A$, the simulation results
in a convergent algorithmic run of $A'$ consisting of the same number 
of algorithmic steps.
\begin{theorem}
\label{theorem-alg-equiv}
Let $A = (\Sigma,\vecG,\cI)$ and $A' = (\Sigma',\vecG',\cI')$ be 
concurrent proto-algorithms, where 
$\cI = (D,\Din,\Dout,I)$, and $\cI' = (D',\Din',\Dout',I')$.
Then $A \asim A'$ only if there exist total functions 
$\funct{\fncI}{\Din}{\Din'}$ and $\funct{\fncO}{\Dout'}{\Dout}$ 
such that: 
\begin{enumerate}
\item[(1)]
for all $\din \in \Din$ and $\dout \in \Dout$,
$(\din,\dout) \in \widehat{A}$ only if 
there exists a $\dout' \in \Dout'$ such that
$(\fncI(\din),\dout') \in \widehat{A'}$ and $\fncO(\dout') =  \dout$;
\item[(2)]
for all $\din \in \dom \widehat{A}$, 
for all convergent $\sigma \in \arun_A(\din)$, 
there exists a convergent $\sigma' \in \arun_{A'}(\fncI(\din))$ with
$\fncO(\outputx_{A'}(\sigma')) = \outputx_A(\sigma)$ such that 
$|\sigma| = |\sigma'|$.
\end{enumerate}
\end{theorem}
\begin{proof}
Because $A \asim A'$, there exists an algorithmic simulation of $A$ by 
$A'$.

Let $R$ be an algorithmic simulation of $A$ by $A'$,
let $\fncI$ be a function from $\Din$ to $\Din'$ such that, 
for all $\din \in \Din$, 
$((\din,\botc,\bot),(\fncI(\din),\botc,\bot)) \in R$, and
let $\fncO$ be a function from $\Dout'$ to $\Dout$ such that, 
for all $\dout' \in \Dout'$, 
$((\bot,\botc,\fncO(\dout')),(\bot,\botc,\dout')) \in R$.
Functions $\fncI$ and $\fncO$ exist by the definition of an algorithmic
simulation.
By Lemma~\ref{lemma-simulation}, 
for all $\din \in \Din$, for all $\sigma \in \arun_A(\din)$,
there exists a $\sigma' \in \arun_{A'}(\fncI(\din))$ such that,
for all $n \in \Natpos$, $(\sigma[n],\sigma'[n]) \in R$.

Let $\din \in \Din$, and
let $\sigma \in \arun_A(\din)$ and $\sigma' \in \arun_{A'}(\fncI(\din))$ 
be such that, for all $n \in \Natpos$, $(\sigma[n],\sigma'[n]) \in R$.
Then from the definition of an algorithmic simulation, it immediately 
follows that, for all $n \in \Natpos$:
\begin{enumerate} 
\item[(a)]
$\sigma[n] \in \Sfin_A$ only if $\sigma'[n] \in \Sfin_{A'}$; 
\item[(b)]
for all $\din \in \Din$, $\sigma[n] = (\din,\botc,\bot)$ only if 
$\sigma'[n] = (\fncI(\din),\botc,\bot)$;
\item[(c)]
for all $\dout \in \Dout$, $\sigma[n] = (\bot,\botc,\dout)$ only if 
there exists a $\dout' \in \Dout'$ such that 
$\sigma'[n] = (\bot,\botc,\dout')$ and $\fncO(\dout') = \dout$.
\end{enumerate}
Now, by the definition of the relation computed by a concurrent 
proto-algorithm, (1) both and (2) follows immediately from(a), (b), 
and~(c).
\qed
\end{proof}
It is easy to see that Theorem~\ref{theorem-alg-equiv} goes through as 
far as (1) and~(2) are concerned if algorithmic simulation is replaced by 
computational simulation.
However, (3) does not go through if algorithmic simulation is replaced by 
computational simulation.

The following theorem tells us how isomorphism, algorithmic equivalence, 
and computational equivalence are related.
\begin{theorem}
\label{theorem-equivs}
Let $A$ and $A'$ be concurrent proto-algorithms.
Then: 
\begin{trivlist}
\item[]
$\qquad$ (1) $\;$ $A \iso A'$ only if $A \aeqv A'$ 
$\qquad$ (2) $\;$ $A \aeqv A'$ only if $A \ceqv A'$.
\end{trivlist}
\end{theorem}
\begin{proof}
Let $A = (\Sigma,\vecG,\cI)$ and $A' = (\Sigma',\vecG',\cI')$ be 
concurrent proto-algo\-rithms, where
$\Sigma = (F,\setf{F},\getf{F},P)$, 
$\Sigma' = (F',\setf{F}',\getf{F}',P')$, 
$\vecG = (G_1,\ldots,G_n)$ with
$G_i = (V_i,E_i,\LBLv_i,\LBLe_i,l_i,r_i)$ ($1 \leq i \leq n$), 
$\vecG' = (G'_1,\ldots,G'_{n'})$ with
$G'_j = (V'_j,E'_j,\LBLv'_j,\LBLe'_j,l'_j,r'_j)$ 
($1 \leq j \leq n'$), 
$\cI = (D,\Din,\Dout,I)$, and $\cI' = (D',\Din',\Dout',I')$.

Part~1.
Because $A \iso A'$, there exist bijections $\bijX$, $\bijV_1$, \ldots, 
$\bijV_n$, $\bijD$, $\bijI$, and $\bijO$ as in the definition of $\iso$.
Let $\bijX$, $\bijV_1$, \ldots, $\bijV_n$, $\bijD$, $\bijI$, and $\bijO$ 
be bijections as in the definition of $\iso$, 
let $\bijXb$, $\bijVb_1$, \ldots, $\bijVb_n$, $\bijDb$, $\bijIb$, and 
$\bijOb$ be the extensions of $\bijX$, $\bijV_1$, \ldots, $\bijV_n$, 
$\bijD$, $\bijI$, and $\bijO$, respectively, with the dummy value $\bot$ such 
that $\bijXb(\bot) = \bot$, $\bijVb_1(\bot) = \bot$, \ldots, 
$\bijVb_n(\bot) = \bot$, $\bijDb(\bot) = \bot$, $\bijIb(\bot) = \bot$, 
and $\bijOb(\bot) = \bot$, and
let $\beta$ be the bijection from $\cS_A$ to $\cS_{A'}$ defined by 
$\beta(\din,((v_1,\ldots,v_n),(d_1,\ldots,d_n),d,i),\dout) = 
 (\bijIb(\din),\,
  ((\bijVb_1(v_1),\ldots,\bijVb_n(v_n)),\,
   (\bijDb(d_1),\ldots,\bijDb(d_n)),\,
   \bijDb(d),\,\bijXb(i)),\,
  \bijOb(\dout))$\,. 
Moreover, let $R = \set{(s,\beta(s)) \where s \in \cS_A}$ and
let $R' = \set{(s,\beta^{-1}(s)) \where s \in \cS_{A'}}$. 
Then $R$ is an algorithmic simulation of $A$ by $A'$ and $R'$ is an 
algorithmic simulation of $A'$ by $A$.
This is easily proved by showing that the conditions from the definition 
of an algorithmic simulation are satisfied for all $(s,s') \in R$ and
for all $(s,s') \in R'$, respectively.
Moreover, it follows immediately from the definitions of $R$ and $R'$ 
that $R' = R^{-1}$.
Hence, $A \aeqv A'$.

Part~2.
Because $A \aeqv A'$, there exists an algorithmic simulation of $A$ 
by $A'$ such that its inverse is an algorithmic simulation of $A'$ by 
$A$. 
Let $R$ be an algorithmic simulation of $A$ by $A'$ such that $R^{-1}$ 
is an algorithmic simulation of $A'$ by $A$.
Then $R$ is also a computational simulation of $A$ by $A'$ and $R^{-1}$ 
is also an computational simulation of $A'$ by $A$.
This is easily proved by showing that the conditions from the definition 
of a computational simulation are satisfied for all $(s,s') \in R$ and
for all $(s,s') \in R'$, respectively.
Hence, $A \ceqv A'$.
\qed
\end{proof}
The opposite implications do not hold in general.
That is, there exist concurrent proto-algorithms $A$ and $A'$ for which 
it does not hold that $A \iso A'$ if $A \aeqv A'$ and there exist 
concurrent proto-algorithms $A$ and $A'$ for which it does not hold 
that $A \aeqv A'$ if $A \ceqv A'$.
In both cases, the construction of a general illustrating example can be
obtained from the construction of a general illustrating example for 
classical proto-algorithms described in~\cite{Mid24a} by applying that
construction to a component of a concurrent proto-algorithm.

The definition of algorithmic equivalence suggests that it is reasonable
to consider the patterns of behaviour expressed by algorithmically 
equivalent concurrent proto-algorithms the same.
This suggests in turn that concurrent algorithms can be considered 
equivalence classes of concurrent proto-algorithms under algorithmic 
equivalence.
The definition of computational equivalence does not suggest that it is 
reasonable to consider the patterns of behaviour expressed by 
computationally equivalent concurrent proto-algorithms the same because 
steps that consist of inspecting a condition are treated as if they do 
not belong to the patterns of behaviour.
The relevance of the computational equivalence relation is that any 
equivalence relation that captures the sameness of the patterns of 
behaviour expressed by concurrent proto-algorithms to a higher degree 
than the algorithmic equivalence relation must be finer than the 
computational equivalence relation.

\section{Concurrency versus Non-determinism in Algorithms}
\label{sect-non-det-concur}

Until now, only concurrent proto-algorithms with deterministic 
components have been considered.
In this section, concurrent proto-algorithms with non-deterministic 
components are also considered because there is an interesting 
connection between concurrency and non-determinism in the setting of
proto-algorithms. 

In order to define the notion of a concurrent proto-algorithm with
non-deterministic components, we have to weaken, in the definition of 
the notion of a concurrent-$\Sigma$-algorithm component graph, the 
outdegree of vertices labeled with a function symbol other than $\fin$
to greater than or equal to 1. 
\begin{udef}
Let $\Sigma = (F,\setf{F},\getf{F},P)$ be an alphabet.
A 
\emph{non-deterministic con\-current-$\Sigma$-algorithm component graph} 
$G$ is a rooted labeled directed graph $(V,E,\LBLv,\LBLe,l,r)$ that 
satisfies the same conditions as a concurrent-$\Sigma$-algorithm 
component graph except that the condition
\begin{quote}
if $l(v) \in \widehat{F} \union \setf{F} \union \getf{F}$, then
$\outdeg(v) = 1$ and, for the unique $v' \in V$ such that 
$(v,v') \in E$, $l((v,v'))$ is undefined
\end{quote}
is replaced by the condition
\begin{quote}
if $l(v) \in \widehat{F} \union \setf{F} \union \getf{F}$, then
$\outdeg(v) \geq 1$ and, for each $v' \in V$ such that 
$(v,v') \in E$, $l((v,v'))$ is undefined.
\end{quote}
\end{udef}
Now, the definition of the notion of a concurrent proto-algorithm with 
non-deterministic components is simple.
\begin{udef}
A \emph{concurrent proto-algorithm} $A$ \emph{with non-deterministic 
components} is a triple $(\Sigma,\vecG,\cI)$, 
where:
\begin{itemize}
\item
$\Sigma$ is an alphabet, called the \emph{alphabet} of $A$;
\item
$\vecG$ is a tuple of non-deterministic concurrent-$\Sigma$-algorithm 
component graphs, 
called the \emph{algorithm component graphs} of $A$;
\item
$\cI$ is a $\Sigma$-interpretation, 
called the \emph{interpretation} of $A$.
\end{itemize}
\end{udef}

It is easy to see that all definitions and results concerning concurrent 
proto-algorithms given in this paper, except 
Corollary~\ref{corollary-classical}, go through for concurrent 
proto-algorithms with non-deterministic components.

A special case of a concurrent proto-algorithm with non-deterministic 
components is a non-deterministic sequential proto-algorithm.
\begin{udef}
A \emph{non-deterministic sequential proto-algorithm} $A$ is a
concurrent proto-algorithm $(\Sigma,\vecG,\cI)$ with non-deterministic 
components where $\Sigma$ is a classical alphabet and $\vecG$ is a tuple 
of one non-deterministic concurrent-$\Sigma$-algorithm component graph.
\end{udef}
In the light of Corollary~\ref{corollary-classical}, the notion of a
non-deterministic sequential proto-algorithm may be viewed as the 
non-deterministic variant of the notion of a classical proto-algorithm. 

The following theorem concerns the connection between concurrent 
proto-algorithms and non-deterministic sequential proto-algorithms.
\begin{theorem}
\label{theorem-non-det-concur}
Let $A$ be a concurrent proto-algorithm.
Then there exists a non-deterministic sequential proto-algorithm $A'$
such that $A \aeqv A'$.
\end{theorem}
\begin{proof}
Let $A = (\Sigma,\vecG,\cI)$ be a concurrent proto-algorithm, where 
$\Sigma = (F,\setf{F},\getf{F},P)$, $\vecG = (G_1,\ldots,G_n)$ with
$G_i = (V_i,E_i,\LBLv_i,\LBLe_i,l_i,r_i)$ (\mbox{$1 \leq i \leq n$}), 
and $\cI = (D,\Din,\Dout,I)$, and
let $j \in \set{1,\ldots,n}$ be such that $G_j$ is the
concurrent-$\Sigma$-algorithm main component graph. 
Then we can construct a non-deterministic sequential proto-algorithm 
$A' = (\Sigma',(G'),\cI')$ from the concurrent proto-algorithm $A$ such 
that $A \aeqv A'$ as described below.
\\*[1.5ex]
$\Sigma'$ is constructed from $\Sigma$ as follows: 
$\Sigma' = (F',\emptyset,\emptyset,P')$, where:
\begin{list}{}{\setlength{\leftmargin}{1.5em}}
\item 
$F' =
 (\set{f_i \where
       f \in \procf{F} \union \setf{F} \union \getf{F} \Land
       i \in \set{1,\ldots,n}} \union
  \set{\ini,\fin}$;
\item
$P' =
  \set{p_i \where p \in P \Land i \in \set{1,\ldots,n}}$.
\end{list}
$\cI'$ is constructed from $\cI$ as follows: 
$\cI' = (D^{n+1},\Din,\Dout,I')$, where:
\begin{list}{}{\setlength{\leftmargin}{1.5em}}
\item 
$I'(f)(d_1,\ldots,d_{n+1}) \, = 
 (d_1,\ldots,d_{j-1},I(f)(d_j),d_{j+1},\ldots,d_{n+1})$
if $f \in \set{\ini,\fin}$;
\item
$I'(f_i)(d_1,\ldots,d_{n+1}) =
 (d_1,\ldots,d_{i-1},I(f)(d_i),d_{i+1},\ldots,d_{n+1})$
if $f \in \procf{F}$;
\item
$I'(f_i)(d_1,\ldots,d_{n+1}) = (d_1,\ldots,d_n,I(f)(d_i,d_{n+1}))$
if $f \in \setf{F}$;
\item
$I'(f_i)(d_1,\ldots,d_{n+1}) = 
 (d_1,\ldots,d_{i-1},I(f)(d_i,d_{n+1}),d_{i+1},\ldots,d_{n+1})$
if $f \in \getf{F}$;
\item
$I'(p_i)(d_1,\ldots,d_{n+1}) = I(p)(d_i)$
if $p \in P$.
\end{list}
$G'$ is the restriction of a rooted labeled directed graph $G$ to the
vertices and edges reachable from its root.
The graph $G$ is constructed from $\vecG$ as follows: 
$G = (V,E,\LBLv,\LBLe,l,r)$, where:
\begin{list}{}{\setlength{\leftmargin}{1.5em}}
\item 
$V = (V_1 \x \ldots \x V_n) \x \set{1,\ldots,n}$;
\item
$E$ is such that:
\begin{list}{}{\setlength{\leftmargin}{1.5em}}
\item 
$(((v_1,\ldots,v_n),i),
  ((v_1,\ldots,v_{i-1},v',v_{i+1},\ldots,v_n),i')) \in E$
\\ \hspace*{1em}
if $(v_i,v') \in E_i$, 
$l(v_i) \in F \union \setf{F} \union \getf{F}$, 
$i' \in \set{1,\ldots,n}$, and $\outdeg(v_{i'}) > 0$;
\item 
$(((v_1,\ldots,v_n),i),
  ((v_1,\ldots,v_{i-1},v',v_{i+1},\ldots,v_n),i))\; \in E$
\\ \hspace*{1em}
if $(v_i,v') \in E_i$ and $l(v_i) \in P$;
\end{list}
\item
$\LBLv = F' \union P'$;
\item
$\LBLe = \set{0,1}$;
\item
$l$ is such that
\begin{list}{}{\setlength{\leftmargin}{1.5em}}
\item 
$l(((v_1,\ldots,v_n),i)) = f$\, if $l_i(v_i) = f$ and 
$f \in \set{\ini,\fin}$;
\item 
$l(((v_1,\ldots,v_n),i)) = f_i$ if $l_i(v_i) = f$ and 
$f \in \procf{F} \union \setf{F} \union \getf{F}$;
\item 
$l(((v_1,\ldots,v_n),i)) = p_i$ if $l_i(v_i) = p$ and $p \in P$;
\item
$l((((v_1,\ldots,v_n),i),((v'_1,\ldots,v'_n),i'))) = l_i((v_i,v'_i))$
if $l_i(v_i) \in P$;
\end{list}
\item
$r = ((r_1,\ldots,r_n),j)$.
\end{list}

The non-deterministic sequential proto-algorithm $A'$ is constructed in 
such a way that there exists a bijection from $\cS_A$ to $\cS_{A'}$ and 
that this bijection, say $\beta$, is such that, for all 
$s,s' \in \cS_A$, $s \in \astep_A(s')$ iff 
$\beta(s) \in \astep_{A'}(\beta(s'))$.
From this, it follows immediately that the relations
$R =  \set{(s,\beta(s)) \where s \in \cS_A}$ and 
$R' = \set{(\beta(s),s) \where s \in \cS_A}$ witness
algorithmic equivalence of $A$ and $A'$.
\qed
\end{proof}
Because concurrent algorithms are expected to be equivalence classes of 
concurrent proto-algorithms under algorithmic equivalence,
Theorem~\ref{theorem-non-det-concur} suggests that each concurrent 
proto-algorithm represents an algorithm that can as well be represented 
by a non-deterministic sequential proto-algorithm.

\section{Concluding Remarks}
\label{sect-conclusions}

Based on the classical informal notion of an algorithm, the notion of a 
classical proto-algorithm has been introduced in~\cite{Mid24a}.
The notion of a concurrent proto-algorithm introduced in this paper is 
a generalization of that notion. 
Other interesting generalizations of the notion of a classical
proto-algorithm are the notion of an interactive proto-algorithm and 
notions of a parallel proto-algorithm.

The notion of an interactive proto-algorithm has been introduced 
in~\cite{Mid24c}.
To also cover (modulo algorithmic equivalence) the algorithms that are 
usually considered both concurrent and interactive, we can try to 
combine the notion of a concurrent proto-algorithm and the notion of an 
interactive proto-algorithm.
It is to be expected that the resulting notion will be rather complex.
Theorem~\ref{theorem-non-det-concur} and the fact that the notion of an 
interactive proto-algorithm from~\cite{Mid24c} admits of non-determinism 
suggest that we can instead cover those algorithms by the notion of an 
interactive proto-algorithm.

A notion of a parallel proto-algorithm is more concrete than the 
notion of a concurrent proto-algorithm.
This means that the algorithms covered by a notion of a parallel 
proto-algorithm are also covered by the notion of a concurrent
proto-algorithm.
A notion of a parallel proto-algorithm may, for example, be useful in 
studying the time efficiency of concurrent algorithms implemented using 
a particular kind of parallelism.


\bibliographystyle{splncs04}
\bibliography{ALG}

\end{document}